\documentclass{ws-procs9x6}

\usepackage{graphicx,multirow,xspace,amsmath,amssymb}

\newcommand{\snn}{\ensuremath{\sqrt{s_{NN}}}\xspace}
\newcommand{\Ups}{\ensuremath{\Upsilon}\xspace}

\newcommand{\UpsOne}{\ensuremath{{\Upsilon(\text{1S})}}\xspace}
\newcommand{\UpsTwo}{\ensuremath{{\Upsilon(\text{2S})}}\xspace}
\newcommand{\UpsThree}{\ensuremath{{\Upsilon(\text{3S})}}\xspace}
\newcommand{\UpsExc}{\ensuremath{{\Upsilon(\text{2S+3S})}}\xspace}
\newcommand{\UpsAll}{\ensuremath{{\Upsilon(\text{1S+2S+3S})}}\xspace}
\newcommand{\Jpsi}{\ensuremath{\mathrm{J}/\psi}\xspace}
\newcommand{\pT}{\ensuremath{p_\mathrm{T}}\xspace}

\newcommand{\Raa}{\ensuremath{R_\mathrm{AA}}\xspace}
\newcommand{\Rda}{\ensuremath{R_\mathrm{dAu}}\xspace}
\newcommand{\Npart}{\ensuremath{N_\mathrm{part}}\xspace}
\newcommand{\Ncoll}{\ensuremath{N_\mathrm{coll}}\xspace}

\begin{document}

\title{Production of Quarkonia at RHIC}
%\footnote{\uppercase{T}his work is supported by the 
%\uppercase{H}ungarian \uppercase{OTKA} grant \uppercase{NK73143} and
%by \uppercase{HAESF}, the \uppercase{H}ungarian-\uppercase{A}merican 
%\uppercase{E}nterprise \uppercase{S}cholarship \uppercase{F}und.}}

\author{R\'obert V\'ertesi\footnote{E-mail: vertesi.robert@wigner.mta.hu}}

\address{Wigner Research Centre for Physics of the Hungarian Academy of Sciences RMI, H-1525 Budapest 114, P.O.Box 49, Hungary}
\address{Nuclear Physics Institute of the Academy of Sciences of the Czech Republic, \v{R}e\v{z} 130, 25068 Husinec-\v{R}e\v{z}, Czech Republic}

\maketitle

\abstracts{%
  The production of different quarkonium states provides unique insight to the hot and cold nuclear matter effects in the strongly interacting medium that is formed in high energy heavy ion collisions. While LHC explores the energy frontier, RHIC has a broad physics program to explore the nuclear modification at different energies in a wide range of systems. Some of the most interesting recent results on \Jpsi and \Ups production in p+p, d+Au and A+A collisions from PHENIX and STAR are summarized in this work.}


\section{Introduction}

In the first decade of its operation, the Relativistic Heavy Ion Collider (RHIC) mostly concentrated on measurements of the more abundantly produced particles such as light mesons. These investigations led to the discovery of a strongly interacting quark-gluon plasma (sQGP), and the understanding of its basic properties\cite{Arsene:2004fa,Back:2004je,Adams:2005dq,Adcox:2004mh,Adare:2009qk}. In more recent years, however, vastly improved luminosity and newly installed detector subsystems allowed for rarer probes, such as comprehensive heavy flavor measurements. These probes provide us with tools to explore the thermal and dynamical properties of the sQGP from new aspects.

It has long been suggested that, due to the screening of the heavy quark potential, the production of heavy quarkonia (charmonium and bottomonium states) is suppressed in heavy ion collisions compared to expectations from p+p collisions. Charmonium suppression was, in fact, anticipated as a key signature of quark-gluon plasma (QGP) formation\cite{Matsui:1986dk}. Moreover, different quarkonium states are expected to dissociate (``melt'') in the QGP at different temperatures due to their different binding energies\cite{Digal:2001iu}. It has therefore been suggested that comprehensive quarkonium measurements can be used as a QGP thermometer\cite{Mocsy:2007jz}.

A nuclear modification that is chiefly governed by melting in the sQGP implies a monotonous increase in the suppression of a given quarkonium state with the beam energy. Recent measurements at RHIC and LHC, however, outline a more complicated situation: the beam energy dependence of \Jpsi production is found to be rather weak, moreover, data at LHC energies show less suppression than at RHIC energies\cite{Arnaldi:2012bg}. Our current understanding on this observation is that several nuclear matter effects influence the resulting yields besides melting\cite{Grandchamp:2005yw,Vogt:2012fba,Arleo:2012rs}. In hot nuclear matter, uncorrelated quark-antiquark pairs can recombine in later stages, thus increasing the overall production. Cold nuclear matter effects can also be significant, such as nuclear shadowing and anti-shadowing, initial state energy loss or co-mover absorption of quarks. There is also a sizeable but largely uncertain contribution to the yields from feed-down processes such as $\chi_c$, $\phi$ and $B$ decays to \Jpsi, and $\chi_b$, \UpsTwo and \UpsThree to \UpsOne. Detailed experimental studies are required to disentangle these various effects.

\section{Experimental approach}

RHIC is an extremely versatile accelerator and detector complex that can be utilized to measure different quarkonium states produced in several collisional systems in a wide range of energy, centrality, momentum and rapidity. Symmetric and asymmetric collisions are studied in p+p, p+Au, d+Au, Cu+Cu, Cu+Au, Au+Au and U+U systems over energies ranging from $\sqrt{s_\mathrm{NN}}$=7.7 to 200 GeV for heavy ions, and up to 510 GeV in the case of (polarized or unpolarized) protons.
Both the PHENIX\cite{Adcox:2003zm} and STAR\cite{Ackermann:2002ad} experiments that currently take data at the RHIC ring have a rich heavy flavor program. STAR is capable of measurements in the dielectron channel in full azimuth at midrapidity ($|\eta|${}$<$0.5) using the Time Projection Chamber for tracking, aided by the Barrel Electromagnetic Calorimeter for particle identification. From 2014 onwards, an outer Muon Telescope Detector (MTD) allows for measurements in the dimuon channel with significantly improved resolution\cite{Ruan:2009ug}. Tracking in PHENIX is done using the Drift and Pad Chambers, while the Electromagnetic Calorimeter serves for mid-rapidity particle identification in the central arm (covering $|\eta|${}$<$0.35 in pseudorapidity and $2 \times \pi/2$ in azimuth). The muons are tracked and identified in the forward arms (1.2$<${}$|\eta|${}$<$2.2).

The most common experimental approach is to measure the nuclear modification factor ($\Raa$) of a given state. It is computed by comparing the corrected number of particles measured in A+A collisions to the yield in p+p collisions scaled by the average number of binary nucleon-nucleon collisions, as
\[
\Raa=\frac{\sigma^{inel}_{pp}}{\sigma^{inel}_{AA}}\frac{1}{\langle\Ncoll\rangle}\frac{d\sigma^{AA} / dy}{d\sigma^{pp} / dy}.
\]
Here $\sigma^{inel}_{AA(pp)}$ is the total
inelastic cross section of the A+A (p+p) collisions,
$d\sigma^{AA(pp)}/dy$ denotes the production cross section of the observed state in A+A (p+p) collisions, and $\langle\Ncoll\rangle$ is the average number of binary nucleon-nucleon collisions from the Glauber model. In the case of quarkonia, where the reconstruction is done via the dielectron (dimuon) decay channel, usually the $B_{ee(\mu\mu)} \times d\sigma^{AA(pp)}/dy$ is specified, where $B_{ee(\mu\mu)}$ is the branching ratio corresponding to the given decay process of the state to an electron-positron ($\mu^+ \mu^-$) pair. A similarly defined \Rda is used for describing the nuclear modification in d+Au collisions.

\section{\Jpsi measurements}

\begin{figure}[t]
\begin{minipage}[t]{0.48\textwidth}
\centering
  \includegraphics[width=0.9\textwidth,height=4.5cm]{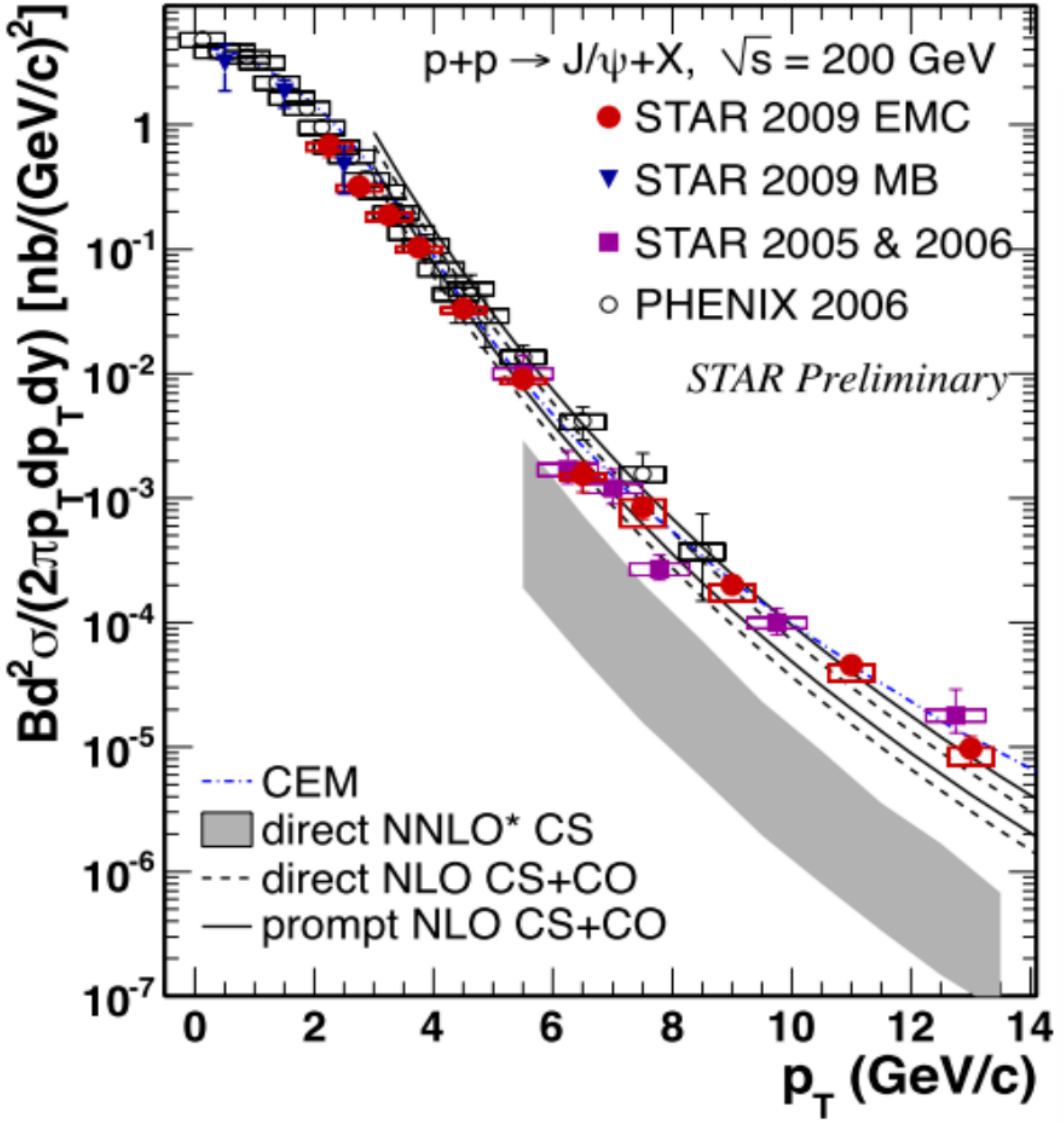}
  \caption{\label{fig:PpJpsiSpectra}
  \pT-spectrum of \Jpsi mesons in p+p collisons at $\sqrt{s_{NN}}$=200 GeV at mid-rapidity measured by PHENIX (open symbols) and STAR (solid symbols), compared to calculations for prompt \Jpsi production.
}
\end{minipage}%
\hspace{0.04\textwidth}%
\begin{minipage}[t]{0.48\textwidth}
  \includegraphics[width=\textwidth,height=4.0cm]{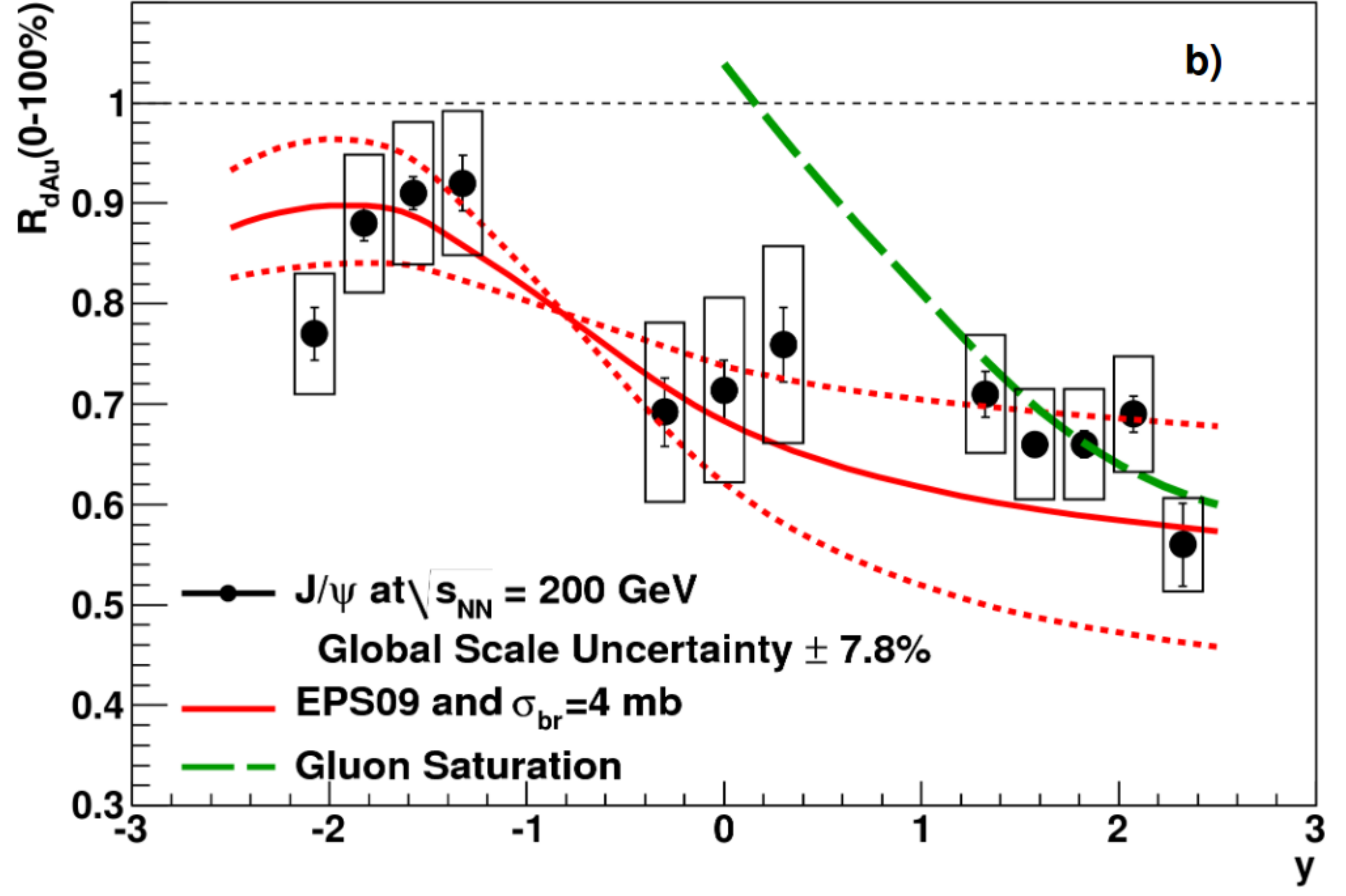}
  \caption{\label{fig:PhnxJpsiRdauY}
  Nuclear modification factor $R_{dAu}$ versus rapidity in d+Au collisions at $\sqrt{s_{NN}}$=200 GeV measured by PHENIX.}
\end{minipage}\hspace{0.04\textwidth}%
\end{figure}

Besides providing reference for nuclear modification in heavier systems, p+p data serve as a benchmark for QCD models. Figure~\ref{fig:PpJpsiSpectra} shows the transverse momentum spectrum of inclusive \Jpsi production at RHIC in p+p collisions at $\sqrt{s_{NN}}$=200 GeV\cite{Adamczyk:2012ey,Kosarzewski:2012zz,Adare:2009js}, compared to several theoretical calculations on prompt \Jpsi production. The prompt color evaporation model\cite{Frawley:2008kk} may provide an acceptable description of the spectrum over the whole transverse monentum range. However, the same models that describe the spectrum face difficulties understanding the \Jpsi polarization\cite{Trzeciak:2013yxr}.

Collisions of d+Au (p+A) systems are generally considered adequate for CNM effect studies because they are too cold for the plasma to form\cite{Adams:2005dq,Adcox:2004mh}. Figure~\ref{fig:PhnxJpsiRdauY} shows the nuclear modification of the \Jpsi mesons in d+Au collisions at \snn{}=200 GeV, measured by PHENIX\cite{Adare:2010fn}. While there is a suppression in the whole rapidity range, the strongest effect is in the gold-going (backward) direction, while it gradually decreases through mid-rapidity and the \Raa is marginally consistent with unity in the deuteron-going (forward) region. This pattern is reasonably explained by the EPS09 model\cite{Eskola:2009uj}, containing nuclear shadowing and final state breakup of the $c\bar{c}$ pair.

\begin{figure}[t]
\begin{minipage}[t]{0.48\textwidth}%
  \centering
  \includegraphics[width=\linewidth]{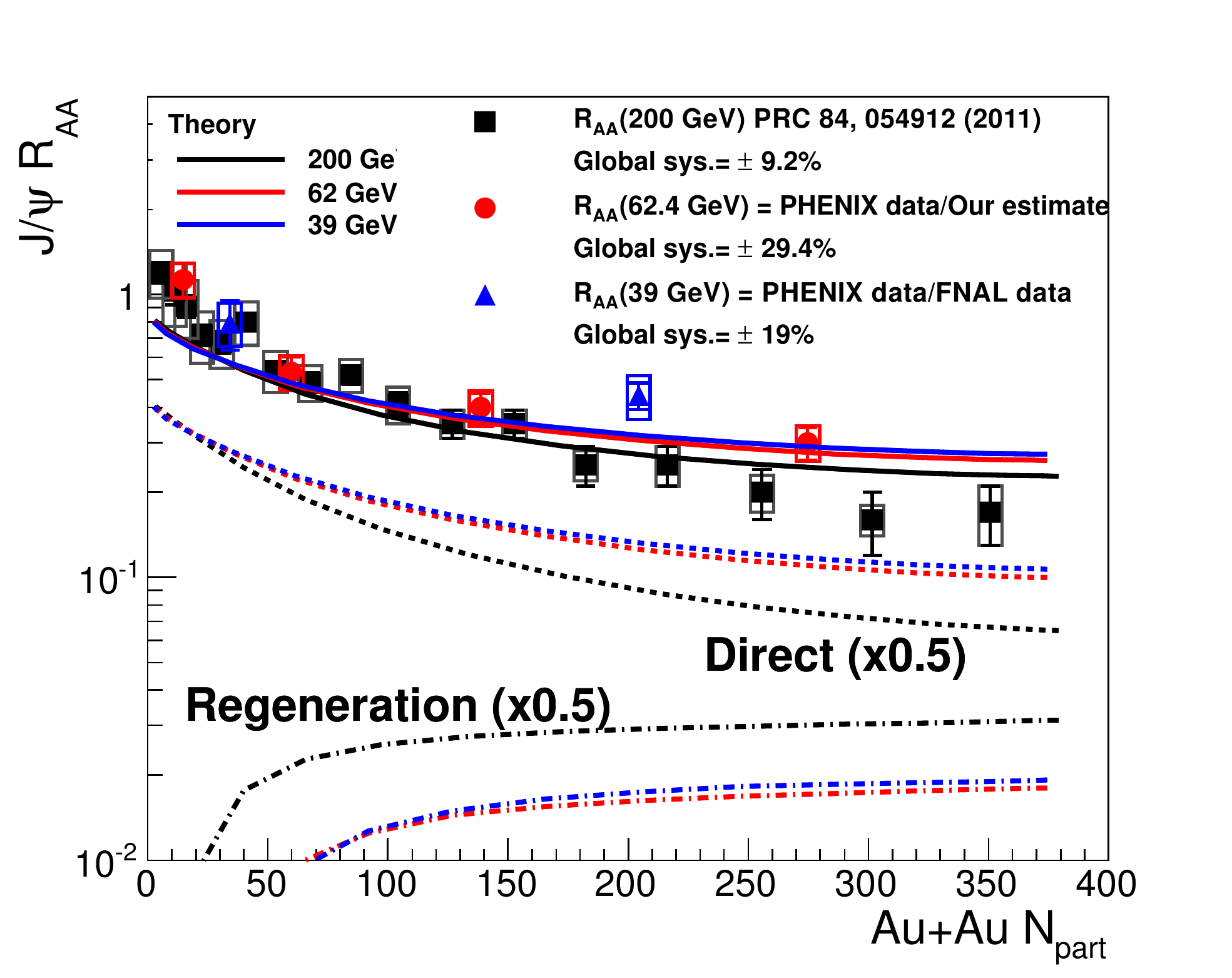}
  \caption{\label{fig:PhenixJpsiBES} 
  PHENIX forward rapidity \Jpsi \Raa vs. \Npart in Au+Au collisions at $\sqrt{s_{NN}}$=39, 62.4 and 200 GeV compared to theoretical calculations.
}
\end{minipage}%
\hspace{0.04\textwidth}%
\begin{minipage}[t]{0.48\textwidth}
%PRC86 064901 (2012)
\centering
  \includegraphics[width=\linewidth,height=4cm]{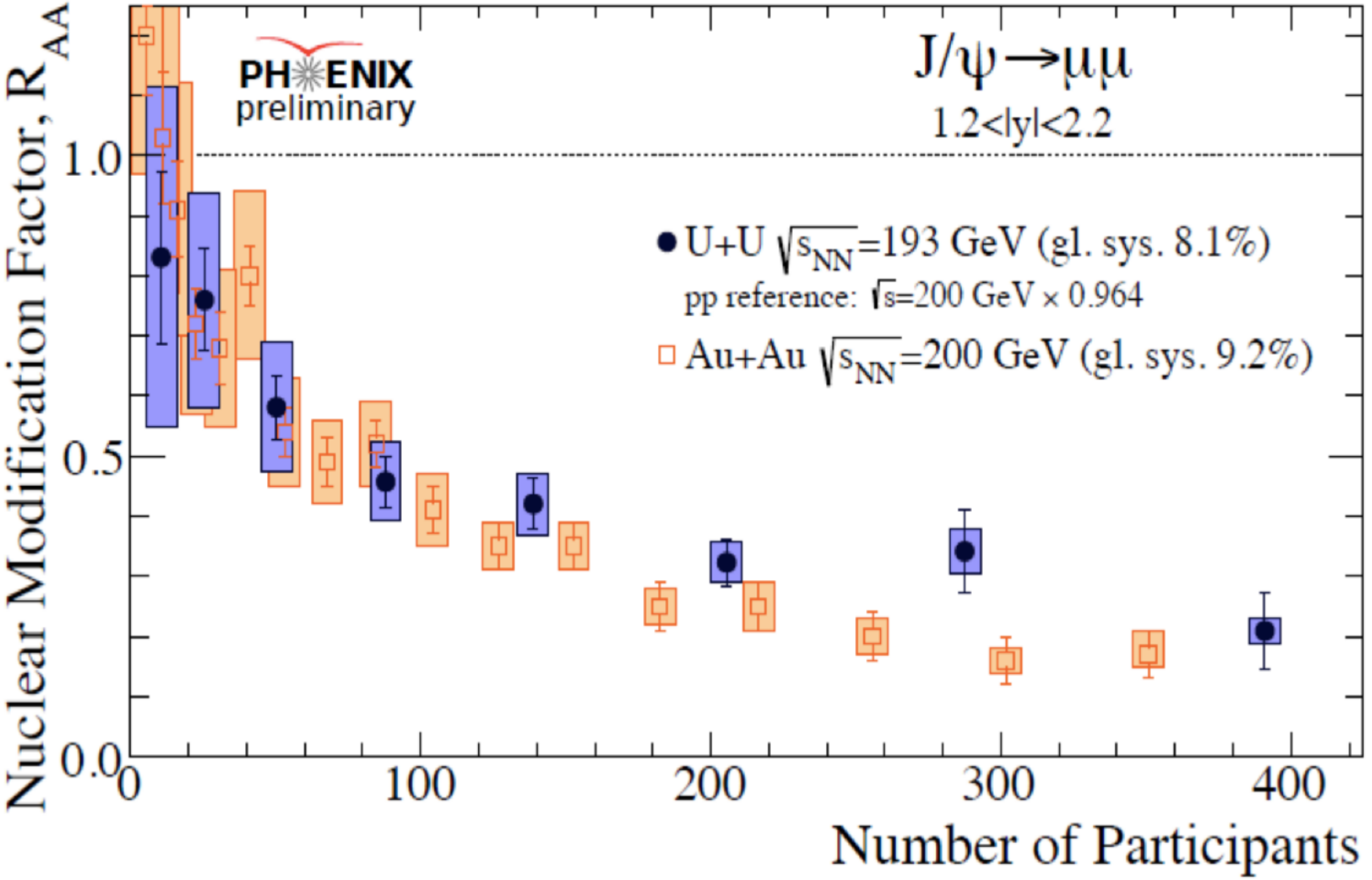}
  \caption{\label{fig:PhenixJpsiAuAuUU} 
  PHENIX forward rapidity \Jpsi \Raa vs. \Npart in Au+Au collisions at $\sqrt{s_{NN}}$=200 GeV (squares) and U+U collisions at  $\sqrt{s_{NN}}$=193 GeV (circles).
}
\end{minipage}%
\end{figure}

Figure~\ref{fig:PhenixJpsiBES} shows the \Jpsi nuclear modification factor at forward rapidities for \snn=39, 62.4 and 200 GeV Au+Au
collisions at PHENIX\cite{Adare:2012wf}, as a function of the number of participant nucleons (\Npart{}). A significant suppression is observed at all energies in mid-peripheral to central collisions, and all three datasets are consistent with each other within uncertainties. STAR data at mid-rapidity exhibits similar trends\cite{Zha:2014nia}. Model predictions that include in-medium dissociation of the \Jpsi as well as later
regeneration from $c\bar{c}$ pairs are consistent with data\cite{Zhao:2010nk}.
The weak dependence of \Raa on collision energy suggests that the different contributions to the \Jpsi nuclear modification largely cancel each other.
Figure~\ref{fig:PhenixJpsiAuAuUU} shows PHENIX forward rapidity measurements of \Raa vs.\ \Npart in U+U and Au+Au systems. The two datasets mostly overlap each other, similarly to other measurements of open and hidden heavy flavor at RHIC\cite{Vertesi:2014tfa}. However, there is a significant discrepancy at higher \Npart values that may be a hint that there is more coalescence in collisions of non-spherical Uranium nuclei.

\begin{figure}[t]
\begin{minipage}[t]{0.48\textwidth}
  \centering
  \includegraphics[width=\linewidth,height=4.5cm]{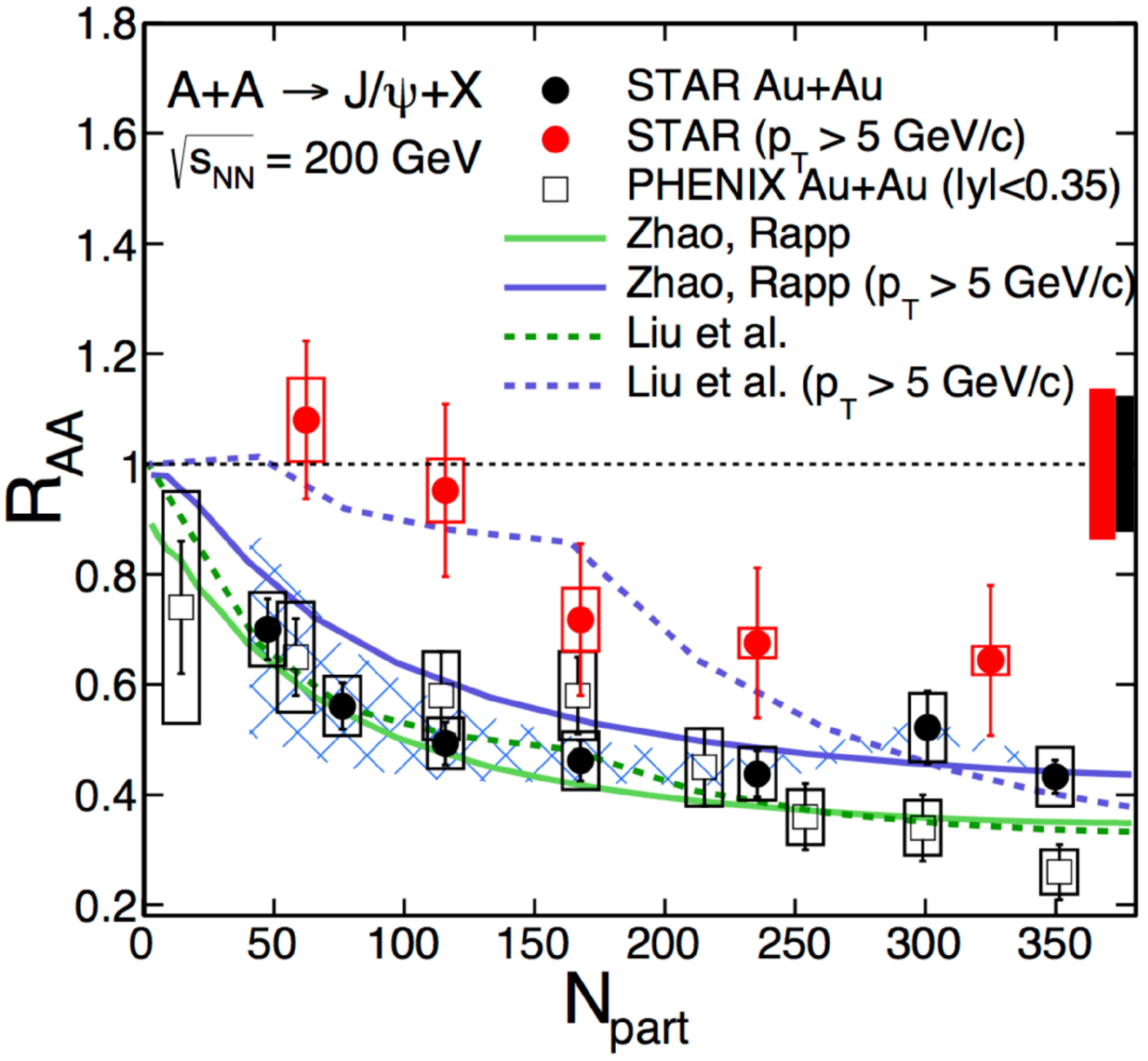}
  \caption{\label{fig:StarHighPtJpsi} 
  STAR and PHENIX minimum bias (black circles and empty squares) \Jpsi vs. \Npart, as well as STAR high-\pT \Jpsi vs. \Npart (shaded circles) in Au+Au collisions at $\sqrt{s_{NN}}$=200 GeV at mid-rapidity, compared to models.
}\end{minipage}%
\hspace{0.04\textwidth}%
\begin{minipage}[t]{0.48\textwidth}
  \centering
  \includegraphics[width=\linewidth,height=4.5cm]{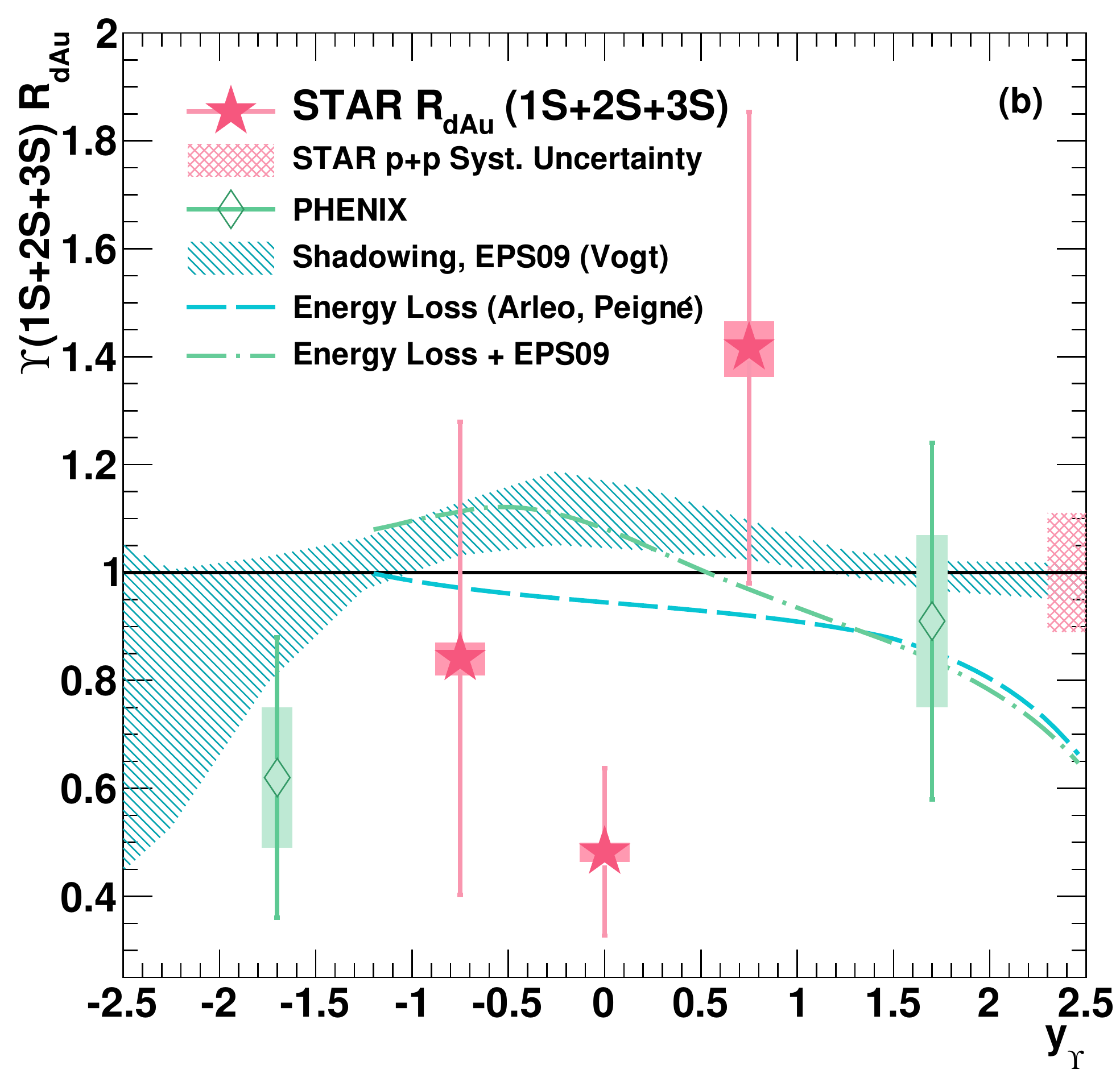}
  \caption{\label{fig:StarUpsRdauY} 
  \UpsAll vs.\ rapidity in $\sqrt{s_{NN}}$=200 GeV d+Au collisions from STAR (stars) and PHENIX (diamonds), compared to theoretical calculations.  
}
\end{minipage}
\end{figure}

We saw that it is difficult to disentangle the sources of nuclear modification, because the \Raa is rather insensitive to beam energy over a wide range and also to the number of nucleons in the chosen heavy ion system. At high momentum, however, the impact of CNM effects and regeneration decreases and hot nuclear modification becomes dominant\cite{Zhao:2010nk}. Figure~\ref{fig:StarHighPtJpsi} shows the nuclear modification factor for \Jpsi mesons with a transverse momentum \pT{}$>$5 GeV/$c$, measured by STAR at mid-rapidity in Au+Au collisions at \snn{}=200 GeV, compared to \Jpsi mesons from the same collisions without momentum selection\cite{Adamczyk:2012ey}. The suppression of high-\pT \Jpsi at higher \Npart values is a clear sign of nuclear modification in the sQGP. 

\section{\Ups production}

Theoretical calculations that include shadowing and/or parton energy loss\cite{Vogt:2012fba,Arleo:2012rs} predict an $\Rda\approx 1$ at mid-rapidity.
While the cross sections for \Ups production in \snn{}=200 GeV p+p collisions\cite{Adamczyk:2013poh} are consistent with NLO pQCD color evaporation model predictions\cite{Vogt:2012fba}, the nuclear modification in d+Au, shown on Figure \ref{fig:StarUpsRdauY}, is stronger than what the above models can explain\cite{Adamczyk:2013poh}. This suggests that CNM effects may not be well understood or that CNM effects alone may not be enough to explain the suppression in the d+Au mid-rapidity bin ($|y|${}$<$0.5). However, the uncertainties on the measurement are still quite high. Evaluating the high-statistics p+Au data taken at RHIC in 2015 is essential to see whether there is an effect beyond model predictions or not.

\begin{figure}[t]
\begin{minipage}[t]{0.48\textwidth}
%PRC86 064901 (2012)
\centering
  \includegraphics[width=\linewidth]{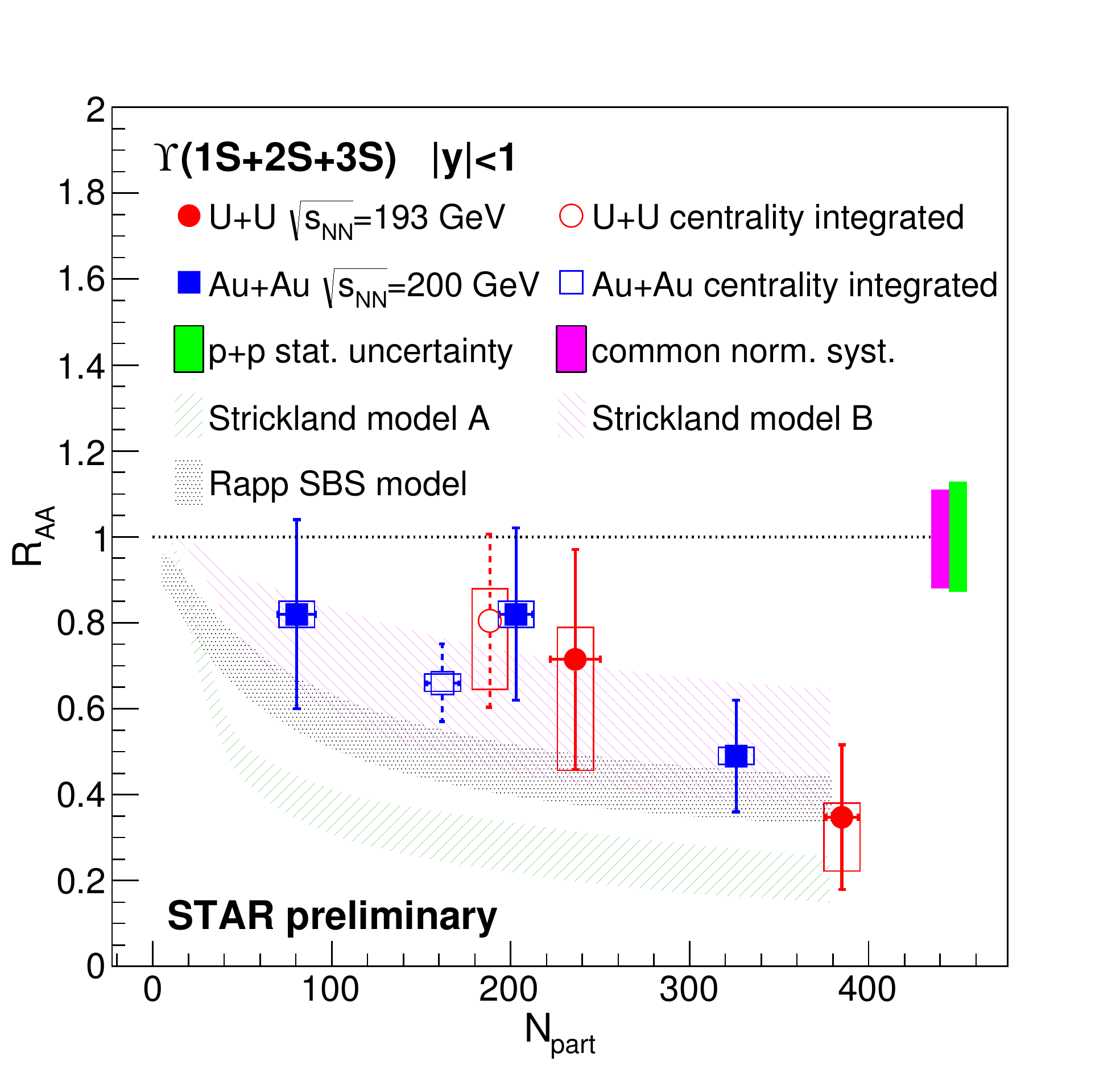}
  \caption{\label{fig:StarUpsRaaNpart}%
  \UpsAll vs. \Npart in $\sqrt{s_{NN}}$=200 GeV Au+Au (squares) and $\sqrt{s_{NN}}$=193 GeV U+U collisions (circles, preliminary), compared to different popular theoretical calculations.
  }
\end{minipage}%
\hspace{0.04\textwidth}%
\begin{minipage}[t]{0.48\textwidth}
  \centering
  \includegraphics[width=\linewidth]{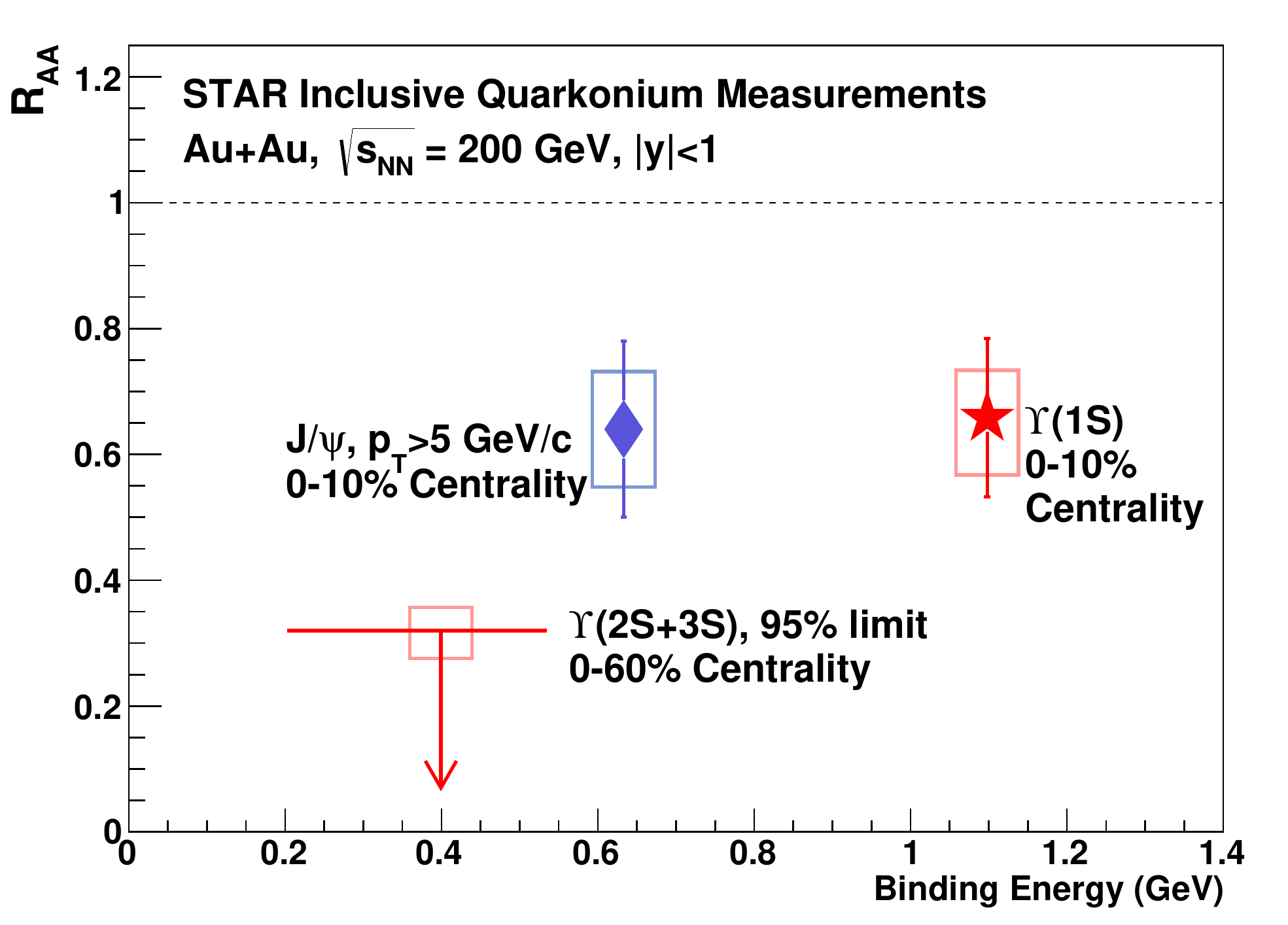}
  \caption{\label{fig:StarUpsRaaEbind} 
  \Raa of central \Ups{}(1S) and the excited states \Ups{}(2S+3S) at 0-60\% centrality are compared to central high-\pT \Jpsi mesons in $\sqrt{s_{NN}}=200$ GeV Au+Au collisions.}
  \end{minipage}
\end{figure}

While the most heavy-ion luminosity in RHIC is collected in Au+Au collisions, central U+U data at $\sqrt{s_{NN}}=193$ GeV are estimated to have a 20\% higher average energy density than that of Au+Au\cite{Kikola:2011zz}, thus allowing for the extension of measurements towards higher \Npart numbers and further tests of the sequential suppression hypothesis. 
Figure \ref{fig:StarUpsRaaNpart} shows the nuclear modification factor of \Ups{}(1S+2S+3S) from Au+Au collisons at $\sqrt{s_{NN}}=200$ GeV as well as from U+U collisions at \snn{}=193 GeV, with respect to \Npart{}. Comparison to theoretical calculations shows the strong $q-\bar{q}$ binding potential scenarios\cite{Strickland:2011aa,Emerick:2011xu} are favored by the data.
\Raa of central \Ups{}(1S) and the excited states \Ups{}(2S+3S) at 0-60\% centrality are compared to central high-\pT \Jpsi mesons in $\sqrt{s_{NN}}=200$ GeV Au+Au collisions\cite{Adamczyk:2012ey,Kosarzewski:2012zz} on Fig.~\ref{fig:StarUpsRaaEbind}. A significant suppression of \Ups states is observed in RHIC central heavy ion collisions\cite{Adamczyk:2013poh,Vertesi:2015fua}. While the \Ups{}(1S) is similarly suppressed to the high-\pT \Jpsi mesons in central Au+Au collisions at $\sqrt{s_{NN}}$=200 GeV, there is a stronger suppression in the case of the excited states. This attests to the sequential melting picture in the presence of a deconfined medium. The suppression pattern seen in the case of the U+U collisions is consistent with the trend marked in Au+Au collisons and extends it towards higher \Npart values.

\section{Summary and outlook}

Recent RHIC measurements show that the presence of a hot medium affects quarkonium production. The significant suppression of high-\pT \Jpsi yields and \UpsOne states in central heavy ion collisions, and an even stronger suppression of \UpsExc states in 0-60\% Au+Au collisions is in accordance with the sequential melting picture. The trends in U+U measurements are generally an extention of those observed in Au+Au to higher number of participant nucleons, although there is some indication in the \Jpsi data that there may be more coalescence in U+U than in Au+Au.
There is no strong dependence of the \Jpsi suppression on the beam energy between $\sqrt{s_{NN}}$=39 and 200 GeV, and the suppression in LHC 2.75 TeV collisions is even weaker than at RHIC, which attests to the role of \Jpsi regeneration. Measurements in d+Au systems show that cold nuclear matter effects are significant in the case of \Jpsi and also may not be negligible in the case of \Ups mesons. 

The developments of PHENIX and STAR, paired with the increased luminosity delivered by RHIC in recent years, will highly increase precision of the measurements in the near future. The first results with STAR MTD have already been presented\cite{Ma:2015xta}. There are ongoing measurements with the inner silicon tracking systems of both experiments, where the prompt and non-prompt \Jpsi contributions will be separated, thus providing means to estimate the fraction of \Jpsi mesons coming from feed-down from $B$ mesons. Besides heavy ion measurements, high luminosity p+Au data recorded in 2015 will help the understanding of cold nuclear matter effects.

\section*{Acknowledgments}

The author would like to thank the Editors of the {\it Exploring Quantum Field Theory} Gribov 85 memorial issue, especially J\'ulia Ny\'\i{}ri for the invitation of this contribution; 
and the organizers of the {\it Joint Wigner-CCNU High-Energy Heavy-Ion Balaton Workshop 2015, Tihany, Hungary} for the invitation to present this material. The work has been supported by the grant 13-20841S of the Czech Science Foundation (GA\v{C}R), as well as the Hungarian OTKA grants NK 101438 and NK 106119.

\end{document}